\documentclass[prl, twocolumn, showpacs, superscriptaddress]{revtex4-2}
\usepackage{amsmath, amssymb}
\usepackage{graphicx}
\renewcommand{\section}[1]{{\par\it #1.---}\ignorespaces}

\begin{document}
\title{Pattern Description of Quantum Phase Transitions in the Transverse Antiferromagnetic Ising Model with a Longitudinal Field}
\author{Yun-Tong Yang}
\affiliation{School of Physical Science and Technology, Lanzhou University, Lanzhou 730000, China}
\affiliation{Lanzhou Center for Theoretical Physics $\&$ Key Laboratory of Theoretical Physics of Gansu Province, Lanzhou University, Lanzhou 730000, China}
\author{Hong-Gang Luo}
\email{luohg@lzu.edu.cn}
\affiliation{School of Physical Science and Technology, Lanzhou University, Lanzhou 730000, China}
\affiliation{Lanzhou Center for Theoretical Physics $\&$ Key Laboratory of Theoretical Physics of Gansu Province, Lanzhou University, Lanzhou 730000, China}
\affiliation{Beijing Computational Science Research Center, Beijing 100084, China}

\begin{abstract}
Despite of simplicity of the transverse antiferromagnetic Ising model with a uniform longitudinal field, its phases and involved quntum phase transitions (QPTs) are nontrivial in comparison to its ferromagnetic counterpart. For example, what is the nature of the mixed-order in such a model and does there exist a disorder phase? Here we use a pattern picture to explore the competitions between the antiferromagnetic Ising interaction, the transverse and longitudinal fields and uncover what kind of pattern takes responsibility of these three competing energy scales, thus determine the possible phases and their QPTs or crossovers. Our results not only unveil rich physics of this paradigmatic model, but also further stimulate quantum simulation by using current available experimental platforms.
\end{abstract}
\maketitle

\section{Introduction}
The Lenz-Ising model introduced initially in order to explain ferromagnetism \cite{Lenz1920, Ising1925, Niss2005} plays a paradigmatic role in many branches of modern physics such as statistical physics \cite{Stanley1987, Kondepudi1998} and condensed matter physics \cite{Chaikin2000, Sachdev2011, Suzuki2013, Dutta2015}. In particular, it is one of central models, interested in quantum simulation \cite{Friedenauer2008, Simon2011, Islam2011, Kim2011, Lewenstein2012, Georgescu2014, Monroe2021, Scholl2021} and quantum annealing \cite{Kadowaki1998, Das2008, Johnson2011, Grass2019}. Intriguing physics involved in such a model originates from the fact that it can exhibit a thermodynamical phase transition from paramagnetic at high temperature to ferromagnetic phases at low temperature in the two-dimensional (2D) case, obtained by Onsager in 1944 \cite{Onsager1944}, in a mathematically exact form. This triggered the development of modern statistical physics, in which the key concepts of universality class and critical scaling in connection with phase transitions have been introduced and described by critical exponents in renormalization group theory \cite{Wilson1975}. Thus the study that finds and classifies new phase transitions keep always as one of scientific frontiers in current condensed matter physics, statistical physics, and related disciplines. 

The thermodynamical phase transition was proved to be absent in the one-dimensional(1D) case, as done early by Ising \cite{Ising1925}, since any thermal fluctuations will break ordered phase in this situation. At zero temperture the thermal excitions go away, and as a consequence, an ordered phase is possible. Fortunately, quantum fluctuations introduced by the transverse field continue to fascinate people due to quantum phase transition (QPT), happening even in such a simple system \cite{Sachdev2011, Suzuki2013, Dutta2015}. It was found that such QPT follows the standard universality calss of classical 2D Ising model. Furthermore, the situation becomes more intriguing once a longitudinal field is applied. In the case of ferromagnetic Ising interaction, the QPT is smeared out but a first-order excited-state QPT, missed in literature for a long time, is found in Ref. \cite{Yang2023a}. In the antiferromagnetic case, the situation is in fact quite unclear since it can not be classified solely by the first-order or continuous QPTs, and thus a somehow awkward concept, namely, mixed-order or hybrid type, has been introduced and applied \cite{Thouless1969, Anderson1969, Bar2014, Sheinman2015, Alert2017, Scholl2021, Lajko2021, Gross2022}.  On the other hand, an additional disordered phase has also been addressed by using fidelity susceptibility method \cite{Bonfim2019}, which was missed by previous studies \cite{Sen2000, Ovchinnikov2003, Campostrini2014, Lin2017, Czischek2018, Yuste2018, Rossini2018, Lajko2021}. 

In the present work we provide a pattern picture \cite{Yang2022a, Yang2022b, Yang2022c} to explore the antiferromagnetic Ising model in the presence of a longitudinal field, in which three characteristic energy scales compete each other: the antiferromagnetic Ising interaction flavors an alternting alignment of up and down spins, and the longitudinal field aligns all spins along the direction of the field while the transverse field introduces quantum fluctuations, driving possible QPTs in the system. Intuitively, two phases must exist: the antiferromagnetic one if the antiferromagnetic Ising interaction dominates over the others and the paramagnetic one if the longitudinal field is strong enough. How about the third case, namely, the transverse field is predominant over the others? We identify these three situations by using the pattern picture, which unveils the intriguing physics involved in the system by the pattern occupancies calculated by the projections of the ground state wavefunction on the patterns. The results show that the third case is characterized by alternative up and down spins but with strong quantum fluctuations, and a hidden QPT/crossover without symmetry breaking is uncovered for the first time. In the following we explore explicitly the system starting from the pattern formulation.

\section{Model and Method}
\begin{figure}[tbp]
\begin{center}
\includegraphics[width = \columnwidth]{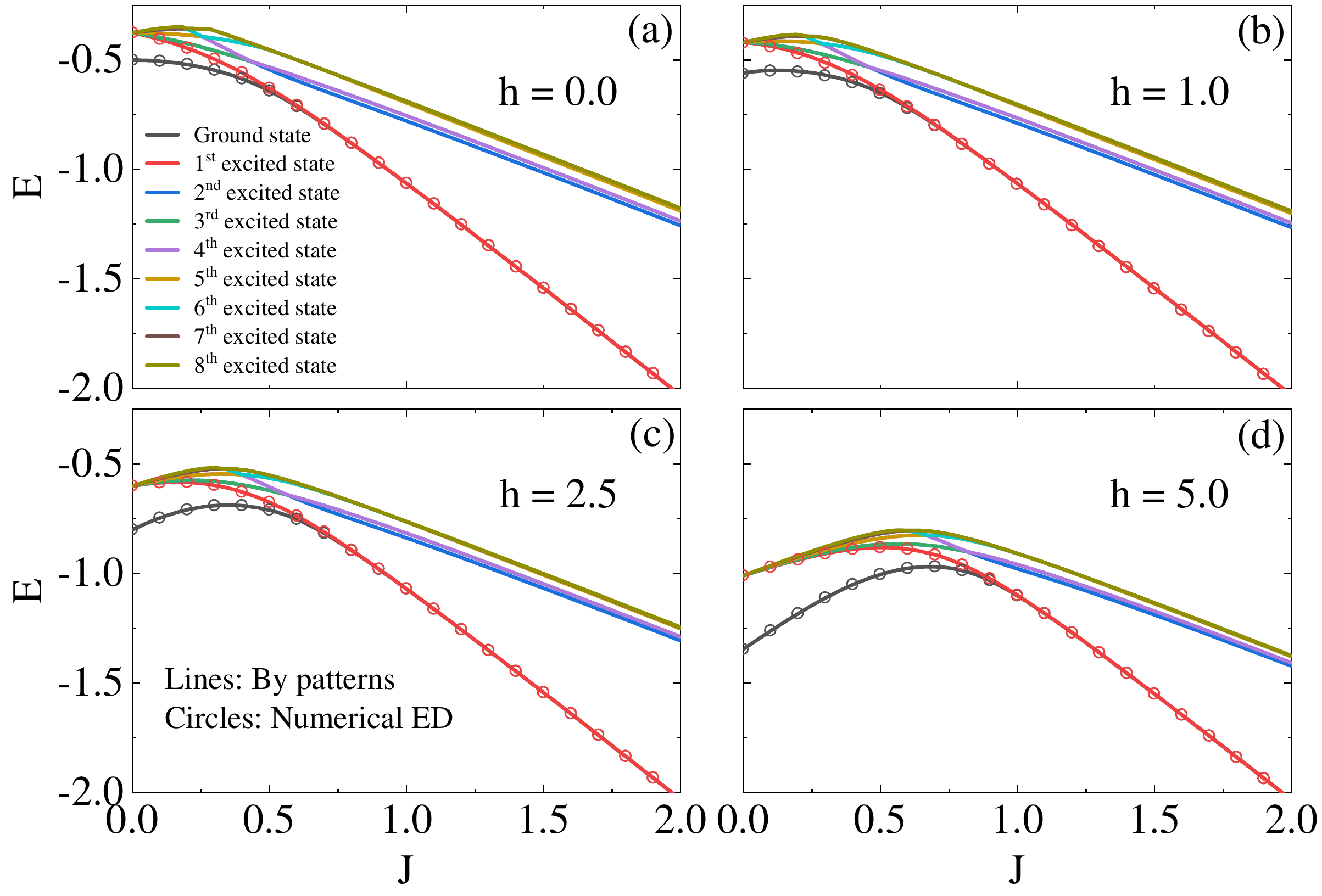}
\caption{The ground state and first eight excited states energies of the transverse antiferromagnetic Ising model in the absence (a) and presence (b, c, d) of longitudinal fields. The circles for the ground and first excited states are obtained by numerical ED, confirming the validity of the present pattern picture.}\label{fig1}
\end{center}
\end{figure}

The transverse antiferromagnetic Ising model Hamiltonian with a longitudial field reads
\begin{equation}
\hat{H}' = J'\sum_{i,\delta}\hat\sigma^z_i \hat\sigma^z_{i+\delta} - h' \sum_i \hat{\sigma}^z_i - g\sum_i \hat\sigma^x_i, \label{Ising0}
\end{equation}
where $J'$, $h'$ and $g$ are assumed to be non-negative here, which denote the antiferromagnetic Ising interaction between two spins representing by Pauli matrix $\hat\sigma$ located at site $i$ and its nearest neighbors $i+\delta$, the longitudinal and transverse fields, respectively. We take the transverse field $g$ as units of energy, and thus quantum fluctuations play an important role in the present context. Eq. (\ref{Ising0}) is rewritten as
\begin{eqnarray}
&& \hat{H}'= \frac{g}{2} \hat{H}, \; \hat{H} = \sum_{i} \hat{H}_i\label{Ising1a} \\
&& \hat{H}_i = -2\hat{\sigma}^x_i - 2h \hat{\sigma}^z_i + J\sum_\delta \left(\hat\sigma^z_i \hat\sigma^z_{i+\delta} + \hat\sigma^z_{i+\delta} \hat\sigma^z_i\right),\label{Ising1b}
\end{eqnarray}
where $J = J'/g$ and $h = h'/g$. For simplicity, we limit ourselves to the 1D case, though it is straightforward to extend to the high-dimensional situations, whose physics will be discussed in future. For a chain with size $L$ under periodic boundary condition (PBC) (i.e.,$\hat{\sigma}^z_{L+1} = \hat{\sigma}^z_1$), $\hat{H}$ can be reformulated as a $3L \times 3L$ matrix in a lattice operator space as follows
\begin{eqnarray}
&&\hat{H} = \left(
\begin{array}{cccccccccc}
\hat{\sigma}^x_1 & -i\hat{\sigma}^y_1& \hat{\sigma}^z_1& \hat{\sigma}^x_2& -i\hat{\sigma}^y_2& \hat{\sigma}^z_2& \cdots& \hat{\sigma}^x_L& -i\hat{\sigma}^y_L& \hat{\sigma}^z_L
\end{array}
\right)\nonumber\\
&&\hspace{0.5cm}\times
\left(
\begin{array}{cccccccccc}
0 & h &0 &0 &0 &0 &\cdots &0 &0 &0 \\
h & 0 &-1 &0 &0 &0 &\cdots &0 &0 &0 \\
0 & -1 &0 &0 &0 &J &\cdots &0 &0 &J \\
0 & 0 &0 &0 &h &0 &\cdots &0 &0 &0 \\
0 & 0 &0 &h &0 &-1 &\cdots &0 &0 &0 \\
0 & 0 &J &0 &-1 &0 &\cdots &0 &0 &0 \\
\vdots & \vdots &\vdots &\vdots &\vdots &\vdots &\ddots &\vdots &\vdots &\vdots \\
0 & 0 &0 &0 &0 &0 &\cdots &0 &h &0 \\
0 & 0 &0 &0 &0 &0 &\cdots &h &0 &-1 \\
0 & 0 &J &0 &0 &0 &\cdots &0 &-1 &0 
\end{array}
\right)\nonumber\\
&&\hspace{0.5cm}\times
\left(
\begin{array}{cccccccccc}
\hat{\sigma}^x_1 & i\hat{\sigma}^y_1 & \hat{\sigma}^z_1 & \hat{\sigma}^x_2 & i\hat{\sigma}^y_2 &\hat{\sigma}^z_2 &\cdots & \hat{\sigma}^x_L & i\hat{\sigma}^y_L & \hat{\sigma}^z_L
\end{array}
\right)^T,\label{Ising2}
\end{eqnarray}
where the identities $\hat{\sigma}^x \hat{\sigma}^y = i \hat{\sigma}^z $ and $\hat{\sigma}^y \hat{\sigma}^z = i \hat{\sigma}^x$ have been used for each site $i$ and the superscript $T$ denotes transpose of the operator vector. The matrix in Eq. (\ref{Ising2}) can be diagonalized to obtain eigenvalues and corresponding eigenfunctions $\{\lambda_n, u_n\} (n = 1, 2, \cdots, 3L)$. As in Refs. \cite{Yang2022a, Yang2022b, Yang2022c, Yang2023a}, we call them patterns marked by $\lambda_n$. With these patterns at hand, $\hat{H}$ is reformulated as
\begin{equation}
\hat{H} = \sum_{n=1}^{3L} \lambda_n \hat{A}^\dagger_n \hat{A}_n, \label{Ising3a} 
\end{equation}
where each pattern $\lambda_n$ composes of one-body operators
\begin{equation}
\hat{A}_n = \sum_{i=1}^{L}\left[u_{n,3i-2} \hat{\sigma}^x_{i}+u_{n,3i-1} (i\hat{\sigma}^y_{i}) + u_{n,3i} \hat{\sigma}^z_{i}\right]. \label{Ising3b}
\end{equation}
Similar to Ref. \cite{Yang2023a}, the validity of Eq. (\ref{Ising3a}) can be confirmed by inserting into a complete basis $|\{\sigma^z_i\}\rangle (i = 1, 2,\cdots, L)$ with $\hat{\sigma}^z_i |\{\sigma^z_i\}\rangle = \pm_i(\uparrow,\downarrow) |\{\sigma^z_i\}\rangle$, as shown in Fig. \ref{fig1} for the ground and first excited states energies obtained by the pattern formulation (lines) and numerical exact diagonalization (ED) (cycles) for several longitudinal fields $h = 0.0, 1.0, 2.5$ and $5.0$ with $L=8$ we take here and hereafter.  

Besides the ground ans first excited states, the other seven excited states are also displayed in Fig. \ref{fig1} in order to provide a complete information about the low-lying states and their dependences on the field applied. For $h = 0.0$ it is very clear that there exists a second-order QPT (see, e.g. Ref.\cite{Yang2022c}) at $J \sim 0.5$. Increasing $J$, the whole structure of energy levels keeps unchanged except for two important points: one is the transition point moves to more large $J$ and the other is each energy level goes rise up to certain $J$, then behaves somehow like the case of $h=0.0$. More larger $h$ is, more visible these two features are. What happens in the presence of a finite longitudinal field? Intuitively, a finite $h$ aligns all spins along the direction of the field, which is obviously paramagnetic phase.  With increasing $J$, the paramagnetic phase is suppressed gradually, but a complete antiferromagnetic order is not built up at once, which needs a more large $J$. Thus a second QPT/crossover should be expected. In the following we study in detail these two QPTs/crossover and identify explicitly these three phases by explicit patterns. 

\begin{figure}[tbp]
\begin{center}
\includegraphics[width = \columnwidth]{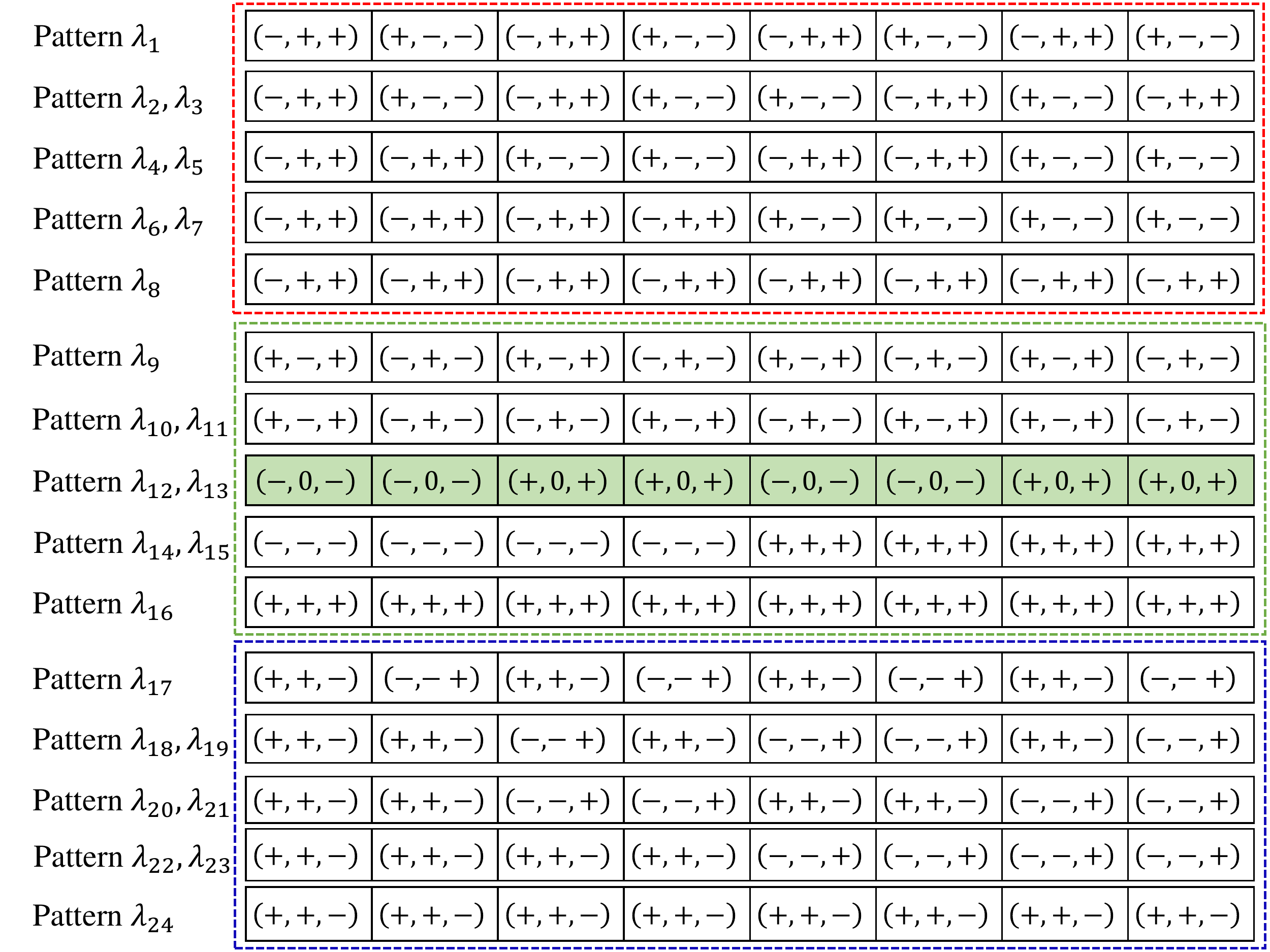}
\caption{The patterns and their relative phases obtained by the first diagonalization, marked by the single-body operators $\hat{A}_n = \sum_{i=1}^L \left[u_{n,3i-2} \hat{\sigma}^x_i+u_{n,3i-1} (i\hat{\sigma}^y_i) + u_{n,3i}\hat{\sigma}^z_i\right]$ with $(\pm,\pm,\pm)$ denoting the signs of $(u_{n,3i-2},u_{n,3i-1},u_{n,3i})$ for the 1D tranverse antiferromagnetic Ising model with $L=8$ under PBC. All patterns are divided into three groups marked by dashed red, dashed green, and dashed blue frames, respectively. For each pattern, a phase factor $e^{i\pi}$ is free, not affecting the relative signs within and between patterns. Marked patterns' energy is zero, dividing all patterns into positive and negative ones.}\label{fig2}
\end{center}
\end{figure}

\begin{figure}[tbp]
\begin{center}
\includegraphics[width = \columnwidth]{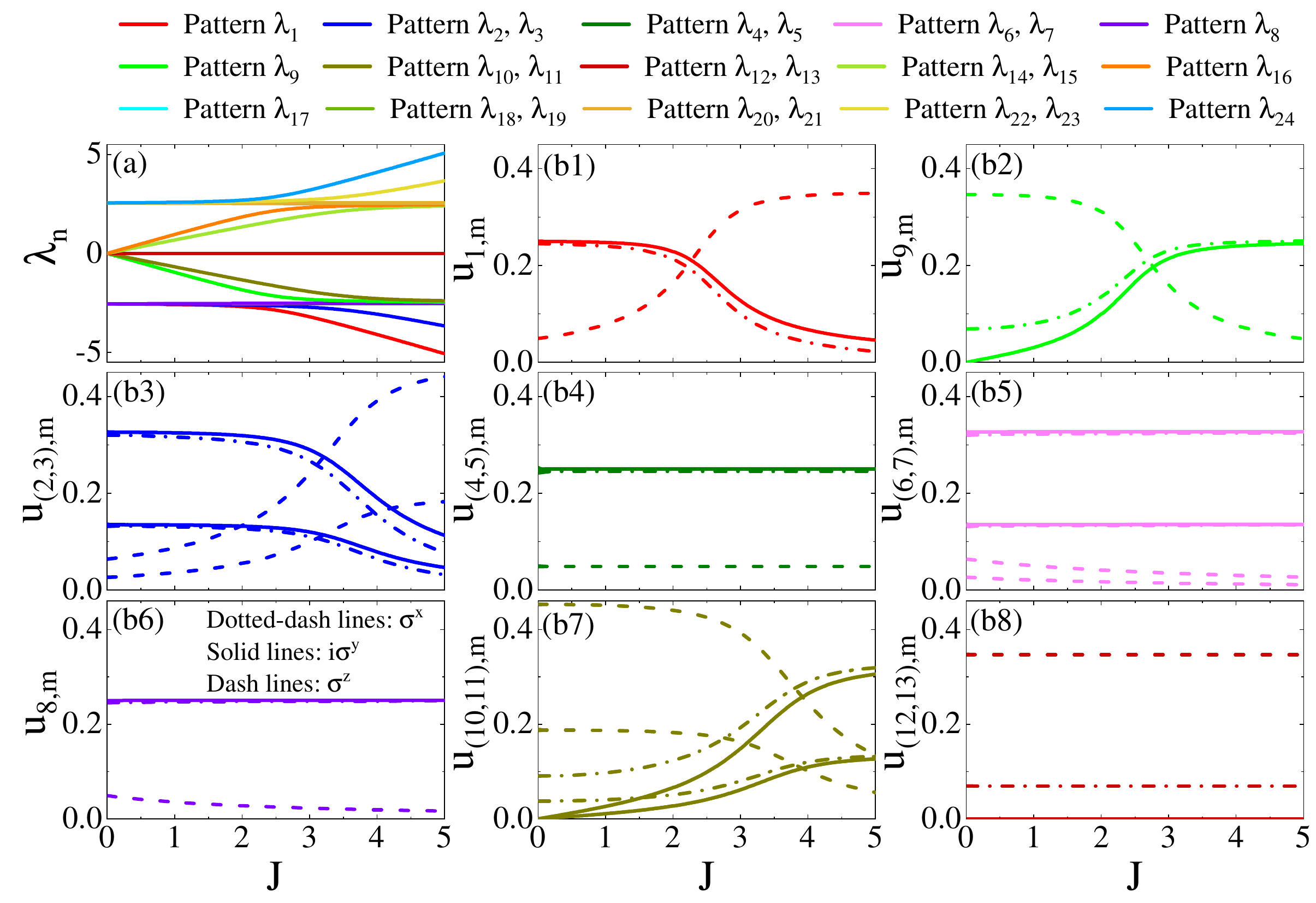}
\caption{(a) The eigenvalues and their eigenfunctions [(b1)-(b8)] of patterns as functions of $J$ and $h = 5.0$ is fixed. It is noted that $\lambda_n = - \lambda_{3L-n+1}$ and $u_{n, m} = - u_{3L-n+1, m}$, and $m$ denotes the spin components. Thus the eigenfunctions from $u_{14,m}$ to $u_{24,m}$ are not shown.} \label{fig3}
\end{center}
\end{figure}
\section{Patterns Properties} 
Figures \ref{fig2} and \ref{fig3} show the patterns and their eigenfunctions dependent of $J$. According to behaviors of the eigenvalues [Fig. \ref{fig3} (a)], all patterns can be divided into three groups, which are degenerate for each group at $J=0$ and the degeneracy is partly lifted with increasing $J$. Their eigenfunctions are smoothly dependent or independent of $J$, showing no any singular behaviors. The signs of patterns show several characteristic features: (i) except for the patterns $\lambda_{12,13}$ with zero pattern eigenvalues, the patterns with negative eigenvalues have the opposite signs for the coefficients of $\hat{\sigma}_i^x$ and $i\hat{\sigma}_i^y$, namely, they are out-of-phase; (ii) oppositely, they are in-phase for the patterns with positive eigenvalues. These two features are the same as its ferromagnetic counterpart \cite{Yang2023a}; (iii) the distinguish between different patterns in each group is the signs of the coefficients of $\hat{\sigma}_i^z$ and their orders, which correspond to different domains or kinks contained in the patterns; (iv) the patterns $\lambda_1$ and $\lambda_9$ exhibit themselves as antiferromagnetic-order-like (the coefficients of $\hat{\sigma}_i^z$ are out-of-phase) and $\lambda_{16}$ looks like ferromagnetic-order-like (the coefficients of $\hat{\sigma}_i^z$ are in-phase). It is necessary to point out that the nature of the patterns, specially the patterns $\lambda_{1}, \lambda_{9}$ and $\lambda_{16}$, remains unchanged even increasing the size of system. This is the reason that the physics of QPTs can be essentially captured even for small lattice size. The finite size effect does not change qualitatively the physics discussed here but affect quantitatively the phase boundary in the phase diagram, which is not discussed here.

\begin{figure}[tbp]
\begin{center}
\includegraphics[width = \columnwidth]{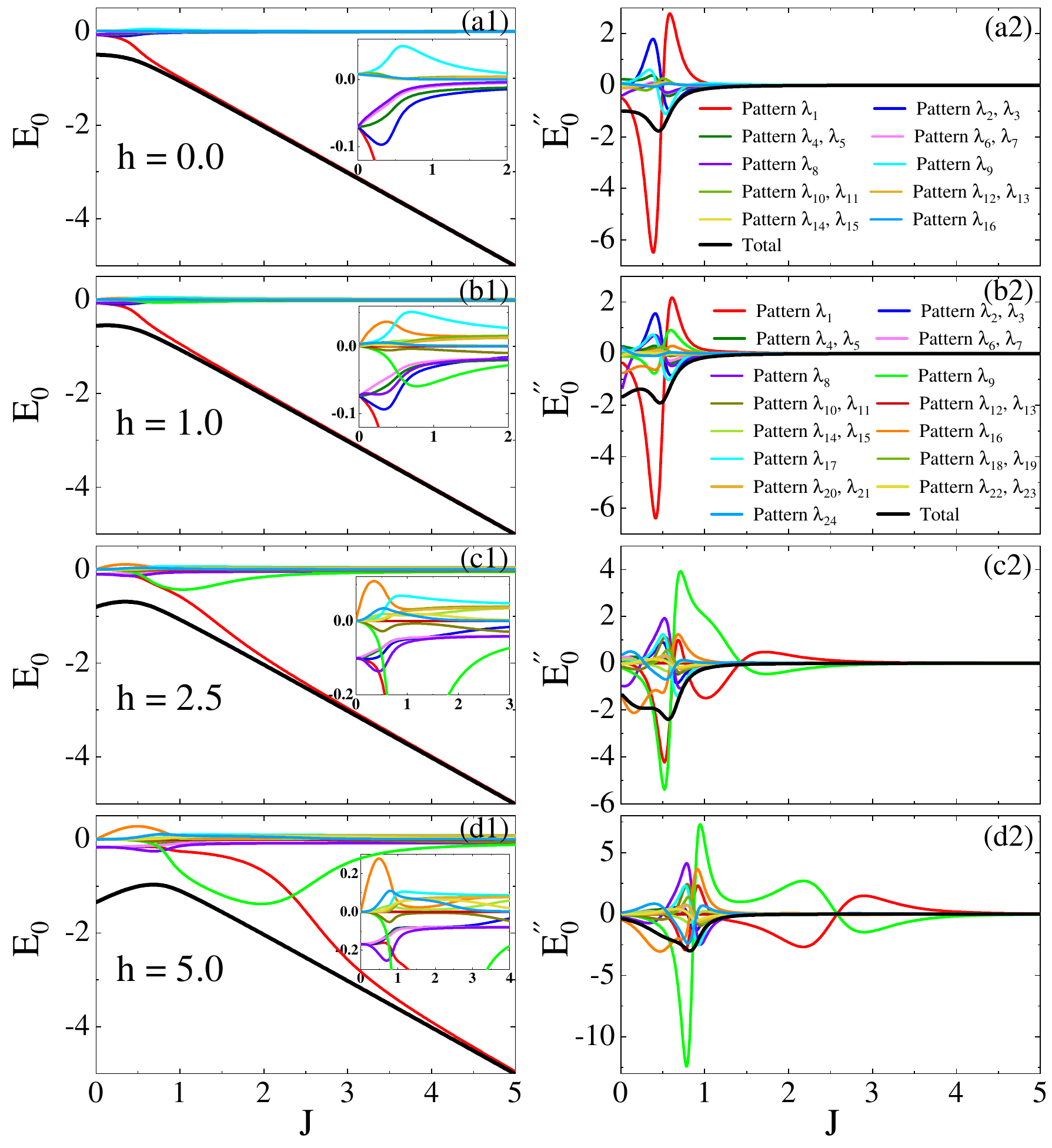}
\caption{(a1)-(d1) The ground state energies (thick solid lines) and their pattern components (thin colored solid lines) as functions of $J$ for $h = 0.0, 1.0, 2.5$ and $5.0$, respectively. The right-upper insets in corresponding plots give an enlarged view of the pattern components. (a2)-(d2) The second derivatives of the energy components of patterns.}\label{fig4}
\end{center}
\end{figure}

\begin{figure*}[tbp]
\begin{center}
\includegraphics[width = 2\columnwidth]{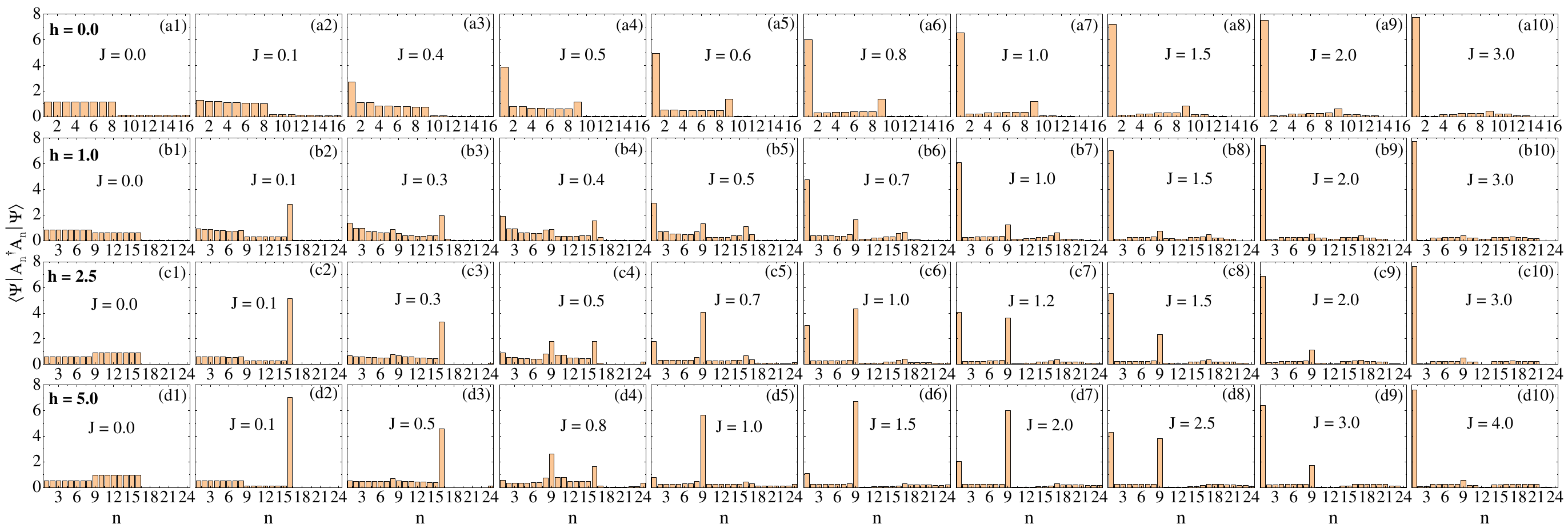}
\caption{Comparison of patterns' occupancy histograms of the ground state for different longitudinal fields $h = 0.0$[(a1)-(a10)], $1.0$[(b1)-(b10)], $2.5$[(c1)-(c10)], and $5.0$[(d1)-(d10)] at different antiferromagnetic Ising interactions $J$'s.}\label{fig5}
\end{center}
\end{figure*}

\section{Two QPTs/Crossovers} 
Figure \ref{fig4} (a1) - (d1) present the ground state energies and their pattern components as functions of the interaction $J$ with $h = 0.0, 1.0, 2.5$ and $5.0$, respectively, as shown by thick black solid lines and thin colored solid lines. In the absence of $h$ shown in Fig. \ref{fig4} (a1) $\&$ (a2), the antiferromagnetic Ising model is essentially equivalent to its ferromagnetic counterpart \cite{Sen2000, Yang2022c}, which is plotted here for comparison. 

Turn to the cases of finite $h$, what role played by the field is apparent: (i) the dominant role of the pattern $\lambda_1$ is delayed obviously once $h$ is beyond the transverse field; (ii) the actions of the patterns $\lambda_{16}$ and $\lambda_9$ are apparently enhanced, in particular, the pattern $\lambda_9$ even plays a dominant role beginning from $h = 2.5$ for a region of $J$[seen, e.g., Fig. \ref{fig4} (d1)]; (iii) comparing Fig. \ref{fig4} (c1) $\&$ (d1) with (a1) $\&$ (b1), the first QPT point remains almost unchanged as $h \leq 1.0$ and it moves toward large $J$ with increasing $h$. At the same time, the role of the pattern $\lambda_1$ is replaced gradually by the pattern $\lambda_9$ at the QPT; (iv) at large enough $J$ the system changes eventully into the antiferromagnetic phase, for example, shown in Fig. \ref{fig4} (d1). However, this transition is only the consequence of the competition between the patterns $\lambda_1$ and $\lambda_9$ and it extends over a wide regime of $J$. In this sense, the second transition is rather a crossover than a QPT; (v) all the above observations can be more clearly seen from the second derivatives of the ground state energies and their pattern components, as shown in Fig. \ref{fig4} (a2)-(d2).

Moving to the histograms of the patterns' occupancies, as shown in Fig. \ref{fig5} for $h = 0.0$ [(a1)-(a10)], $1.0$ [(b1)-(b10)], $2.5$ [(c1)-(c10)], and $5.0$ [(d1)-(d10)] for ten different $J$'s. The patterns' occupancies are calculated by $\langle\Psi|\hat{A}^\dagger_n \hat{A}_n|\Psi\rangle$ where $|\Psi\rangle$ is wavefunction of the ground state. For $h=0.0$, it is completely same as the ferromagnetic Ising model \cite{Yang2022b}, as also mentioned above. 

With finite $h$'s, the situations are completely differnet: in the ferromagnetic case, the ground state QPT disappears \cite{Yang2023a}; but in the antiferromagnetic one, the ground state QPT remains almost unchanged up to $h \sim 1.0$ and the QPT point moves toward large $J$.  From the histograms of pattern components at finite $h$'s, it is noticed that once J is switched on, the system rapidly enters a phase characterized by the pattern $\lambda_{16}$. In this phase, the spins all align along the direction of the longitudinal field. More larger the $h$ is, more stronger the phase is. But it is not a typical or classical ferromagnetic state since there exists strong quantum fluctuations representing by the phase of $\hat{\sigma}^x$ and $i\hat{\sigma}^y$. It is an unstable state since it contributes a positive energy to the ground state of the system. Once increasing $J$, this state is suppressed rapidly, as shown in Fig. \ref{fig5}. At the same time, the pattern $\lambda_9$ becomes increasely dominant. The pattern $\lambda_9$ has a characteristic feature of antiferromagnetic order-like. Likewise, it is not typical or classical antiferromagnetic phase because of quantum fluctuations. In fact, it is a metastable state, as observed in Fig, \ref{fig4}. With further increasing $J$, it is found that the occupancy of the pattern $\lambda_1$ begins to increase gradually and at the same time that of the pattern $\lambda_9$ is suppressed gradually. They become comparable at certain $J$ dependent of $h$ and then the pattern $\lambda_1$ dominates over the others. Thus the system enters the antiferromagnetic phase for a large enough $J \sim h/2$, as expected. 

Some remarks are in order. i) In comparison to the ground state of the ferromgnetic Ising model, that of the antiferromagnetic one exhibits more rich behaviors: the latter experiences three phases ranging from unstable ferromagnetic-like phase, metastable antiferromagnetic-like phase to stable antiferromagnetic phase. The former two phases are not really (anti)ferromagnetic phases in a conventional sense since the quantum fluctuations play an important role here. It should be pointed out that in some literature the unstable ferromgnetic-like phase was labeled by the paramagnetic phase \cite{Novotny1986, Ovchinnikov2003, Zhang2009, Simon2011, Bonfim2019}, the ferromagnetic phase \cite{Lajko2021} and the metastable antiferromagnetic-like phase by the disorder phase \cite{Bonfim2019}. ii) It is understandable why the longitudinal field does not kill the QPTs/crossover in the antiferromagnetic model, obviously different to that in its ferromagnetic counterpart: the effect of the uniform longitudinal field is somehow frustrated by the antiferromgnetic interaction. Thus in a really frustrated system, e.g., the axial next-nearest-neighbor Ising model, a more rich behvior will be expectable, which is left for the future study. iii) The QPT, in fact we prefer to the crossover, between the antiferromagnetic-like phase and the antiferromagnetic one in the large enough $J$ has a wide parameter regime (although one can find $J \sim h/2$ as the crossing point between the energies of the patterns $\lambda_1$ and $\lambda_9$, as shown in Fig. \ref{fig4} at finite $h$), in which the symmetry of the system does not change. Whether this is related to the topological QPT or not remains to be further explored. iv) The present results have been obtained by fixed lattice size. Changing the size does not change the essential physics discussed here, namely, the patterns like $\lambda_1, \lambda_9$ and $\lambda_{16}$ play central roles in understanding the physics of phase transitions, as also pointed out above. Moreover, our results have a realistic significance in quantum simulations such as spin chains in an optical lattices \cite{Simon2011}, trapped ions \cite{Islam2011, Monroe2021}, and Rydberg atom systems \cite{Bernien2017, Keesling2019}, and so on.

\section{Summary}
We use a pattern picture to explore the QPTs/crossover in the transverse antiferromagnetic Ising model with a longitudinal field. The patterns are obtained by diagonalizing the model's Hamiltonian in the operator space, among which three key patterns characterize the phases of the model, namely, the ferromagnetic-like phase, the antiferromagnetic-like phase and the antiferromagnetic phase with a large enough $J \geq h/2$. This physics remains unchanged for different sizes since the in-phase and/or out-of-phase property of the key patterns is independent of the lattice size. Therefore, our results are able to explain why and how the QPTs/crossover happens, which can be readily tested by current quantum simulations. 

\section{Acknowledgments}
The work is partly supported by the National Key Research and Development Program of China (Grant No. 2022YFA1402704) and the programs for NSFC of China (Grant No. 11834005, Grant No. 12247101).

%\bibliography{myrefs}

%%\begin{thebibliography}{99}
%%\bibitem{Greenstein2006} G. Greenstein and A. Zajonc, \textit{The Quantum Challenge: Modern Research on the Founddations of Quantum Mechanics} (Jones and Bartlett Publishers, Sudbury, Massachusetts, 2006).
%%\bibitem{Laloe2019} F. Lalo\"e, \textit{Do We Really Understand Quantum Mechanics?} 2nd Ed. (Cambridge Univ. Press, 2019).
%%\bibitem{Pusey2012} M. F. Pusey, J. Barrett, and T. Rudolph, ``On the reality of the quantum state", Nat. Phys. \textbf{8}, 475-478 (2012).
%%\bibitem{Feynman1989} R. Feynman, \textit{The Feynman Lectures on Physics} (Addison-Wesley, Boston, 1989).
%%\bibitem{Dirac1958} P. A. M. Dirac, \textit{The Principle of Quantum Mechanics} (Oxford University Press, 1958), P74.

%%\end{thebibliography}


\end{document}